\documentclass[a4paper,11pt,usenames,dvipsnames,table]{article}
\pdfoutput=1 

\usepackage{jheppub} 
\usepackage[T1]{fontenc} 
\usepackage{float}
\usepackage[usenames,dvipsnames]{xcolor}
\usepackage{mathtools}
\usepackage[]{mdframed}
\usepackage{amssymb}
\usepackage{amsthm} 
\usepackage{xfrac}

\usepackage{booktabs} 

\title{Holographic entropy inequalities \\pass the majorization test}
\author{Bart{\l}omiej Czech, Yichen Feng, Xianlai Wu, Minjun Xie}
\affiliation{Institute for Advanced Study, Tsinghua University, Beijing 100084, China}

\emailAdd{bartlomiej.czech@gmail.com}
\emailAdd{fyc23@mails.tsinghua.edu.cn}
\emailAdd{wuxl24@mails.tsinghua.edu.cn}
\emailAdd{xiemj21@mails.tsinghua.edu.cn}

\abstract{Quantities computed by minimal cuts, such as entanglement entropies achievable by the Ryu-Takayanagi proposal in the AdS/CFT correspondence, are constrained by linear inequalities. We prove a previously conjectured property of all such constraints: Any $k$ systems on the ``greater-than'' side of the inequality whose overlap is nonempty are subsumed in some $k$ systems on its ``less-than'' side (accounting for multiplicity). This finding adds evidence that the same inequalities also constrain the entropies under time-dependent conditions because it preempts a large class of potential counterexamples. We prove several other properties of holographic entropy inequalities and comment on their relation to quantum erasure correction and the Renormalization Group.
}

\begin{document} 
\maketitle
\flushbottom

\section{Introduction}
\label{sec:introduction}
In quantum mechanics, information is nearly synonymous with quantum entanglement. ``Knowing'' the state of a system is routinely conceptualized as maintaining a memory, which is entangled with the system. It is interesting to identify limits on how this quantum knowledge can be partitioned in systems with multiple components. One universal limit, which is implied directly by the laws of quantum mechanics, is the strong subadditivity of entanglement entropy \cite{ssa1,ssa2}:
\begin{equation}
I(A_1:A_3 | A_2) \equiv S(A_1 A_2) + S(A_2 A_3) - S(A_2) - S(A_1 A_2 A_3) \geq 0
\label{ssa}
\end{equation}
Here $S(A) = -{\rm tr} \rho_A \log \rho_A$ is the von Neumann entropy of subsystem $A$. In quantum field theory the object $\rho_X$ may not be well-defined, but the conditional mutual information $I(A_1 : A_3 | A_2)$ is well-defined and UV-finite so long as subsystems $A_1$ and $A_3$ do not share a common border.

It is also interesting to study these ``fundamental limits on information'' under additional assumptions. In the last two decades, one such assumption has gained prominence. It supposes that the von Neumann entropies of subsystems are computed by minimal cuts through some auxiliary object according to the following formula:
\begin{equation}
S(A) = \min_{{\rm cuts}\,A\, |\, \overline{A}} {\rm Area}({\rm cut})
\label{mincut}
\end{equation}
The primary motivation for considering~(\ref{mincut}) comes from holographic duality or AdS/CFT correspondence \cite{adscft}. In the presence of time reversal symmetry, von Neumann entropies of CFT regions are computed by equation~(\ref{mincut}), a statement known as the Ryu-Takayanagi (RT) proposal \cite{rt1, rt2}. In this context, the cuts are surfaces on the time reversal-symmetric spatial slice of the dual AdS spacetime. Equation (\ref{mincut}) also applies---sometimes up to a gauge subtlety---to many tensor network states as an empirical observation \cite{meraref, holoMERA, HaPPYcode, xiwayne, topoRTN, vijaycharlie}, including random tensor networks \cite{rtn} of arbitrary architecture where it holds as a theorem \cite{pagethm}. Bounds, which are implied by prescription~(\ref{mincut}) are therefore necessary conditions for the emergence of a semiclassical bulk geometry as well as detectors of non-randomness of tensors in tensor network states. In common nomenclature, they are said to bound the {\bf holographic entropy cone} \cite{hec}.

At present, two infinite families and nearly 2000 sporadic inequalities are known to follow from (\ref{mincut}) \cite{monogamy, n5cone,yunfei,longlist,newineqs, facetness}, with infinitely many more awaiting discovery. In this paper, we express these {\bf holographic entropy inequalities} in the notation
\begin{equation}
{\rm LHS} = \sum_{i=1}^L S(X_i) \geq \sum_{j=1}^R S(Y_j) = {\rm RHS},
\label{schematic}
\end{equation}
where regions $X_i$ and $Y_j$ are unions of atomic (indivisible) subsystems $A_p$. Note that all terms in inequality~(\ref{schematic}) come with unit coefficient; this is not restrictive because we allow $X_i = X_{i'}$ or $Y_j = Y_{j'}$. It is useful to have a notation for the number of indivisible subsystems $A_p$ involved in an inequality; we call it $N$ so $1 \leq p \leq N$. As an example, the best known holographic entropy inequality---the monogamy of mutual information \cite{monogamy}---is an $N=3$ inequality:
\begin{equation}
S(A_1 A_2) + S(A_2 A_3) + S(A_1 A_3) \geq S(A_1) + S(A_2) + S(A_3) + S(A_1 A_2 A_3)
\label{mmi}
\end{equation}
Throughout this paper, we denote unions with concatenation, e.g. $A_1 \cup A_2 \equiv A_1 A_2$. 

\subsection{Universal properties of holographic entropy inequalities}
\label{sec:props}
Holographic entropy inequalities~(\ref{schematic}) possess many common properties \cite{kbasis, yunfei, superbalance, sirui, sergiosym, longlist, newineqs, deterministiccontract, nestingnotcontract}, some highly nonobvious. Perhaps the most important such property is that they are completely (or almost completely; see below) UV-finite in field theory. Observe that the UV divergence, which arises from the border of $A_p$ and the unnamed region outside $A_1 A_2 A_3\!\ldots$ contributes to (\ref{schematic}) in proportion to the net number of appearances of $A_p$. The cancellation of such divergences is known as:
\begin{itemize}
\item {\bf Balance.} Let $L_p$ be the number of terms on the LHS of (\ref{schematic}), which contain $A_p$. Define $R_p$ analogously for the RHS. Then $L_p = R_p$ for all $1 \leq p \leq N$. That is, every region appears net zero times in the inequality. 
\end{itemize}
On the other hand, the UV divergence from the common border of $A_p$ and $A_q$ contributes in proportion to the net number of terms, which feature $A_p$ without $A_q$ or vice versa. Combined with balance, the cancellation of such divergence implies:
\begin{itemize}
\item {\bf Superbalance.} Let $L_{pq}$ be the number of terms on the LHS of (\ref{schematic}), which contain the pair $A_p A_q$. Define $R_{pq}$ analogously for the RHS. Then $L_{pq} = R_{pq}$. That is, every pair appears net zero times in the inequality.
\end{itemize}
All holographic entropy inequalities are balanced\footnote{We ignore trivially true but unbalanced inequalities such as $S(A) \geq 0$. In pure states, one may exploit $S(X) = S(\overline{X})$ to rewrite a balanced inequality in an ostensibly unbalanced form. While such rewritings can be useful for exhibiting the full symmetry of an inequality \cite{newineqs}, we do not use them in this paper.} and almost all are superbalanced \cite{superbalance}. The only inequalities, which are not superbalanced, are the subadditivity of entanglement entropy $S(A) + S(B) \geq S(AB)$ and convex combinations of subadditivity with other holographic entropy inequalities. An example of such a convex combination is strong subadditivity~(\ref{ssa}), which is a convex combination of monogamy and subadditivity:
\begin{align}
& S(A_1 A_2) + S(A_2 A_3) + S(A_1 A_3) - S(A_1) - S(A_2) - S(A_3) - S(A_1 A_2 A_3) \nonumber \\
+\, & S(A_1) + S(A_3) - S(A_1 A_3) \geq 0
\end{align}
It is balanced but not superbalanced because the UV-divergence from the common border of $A_1$ and $A_3$ contributes in the second line. 
\medskip

{\bf This paper derives two other properties of holographic entropy inequalities.} They were conjectured in \cite{majorization}, motivated by the expectation that inequalities~(\ref{schematic}) should also hold without the assumption of time reversal symmetry. (Prior evidence for this includes \cite{timedependence, multiconnected3dbulk, HRRandRTconeequals}.) In that circumstance, entropies of boundary subsystems are computed by the Hubeny-Ryu-Takayanagi (HRT) proposal \cite{hrt, maximin}, which involves an additional maximization over spatial slices. Reference~\cite{majorization} observed that when the HRT proposal is used to compute the entropies, regions drawn from the boundary lightcone of a single point saturate inequalities~(\ref{schematic}) in the CFT vacuum. To show that deforming away from the vacuum causes no violation of (\ref{schematic}) would then give significant evidence for the validity of the inequalities in time-dependent settings. One of the properties, which was conjectured in \cite{majorization} and which we prove here, is a mathematical restatement of this desideratum. As such, our results suggest that holographic entropy inequalities are in fact dynamical statements about spacetimes obeying Einstein's equations.

As explained in Reference~\cite{majorization}, the assertion from the previous paragraph holds so long as the holographic entropy inequality passes the following criterion:

\paragraph{Majorization test} 
Given a balanced inequality~(\ref{schematic}), take the following steps:
\begin{enumerate}
\item Construct the $p^{\rm th}$ {\bf null reduction} of the inequality by dropping all terms, which do not contain $A_p$. Balance implies that the null reduction has the same number of terms ($L_p = R_p$) on both sides. As a side note, observe that the null reduction of a superbalanced inequality is balanced because $A_q$ appears on the two sides of the null reduction $L_{pq} = R_{pq}$ times. For example, (\ref{ssa}) is balanced because it is the null reduction on $A_2$ of (\ref{mmi}), which is superbalanced.
\item Form vector $\vec{v}$ (respectively $\vec{z}$) whose components correspond to the summands on the LHS (respectively RHS) in the null-reduced inequality. Each component is the sum of variables $a_q$, which are canonically associated with the atomic regions $A_q$. For example, for (\ref{ssa}) we have $\vec{v} = (a_1+a_2, a_2+a_3)$ and $\vec{z} = (a_2, a_1+a_2+a_3)$.
\item Optionally, we may drop the variable associated with the region on which the null reduction was carried out. By construction, this variable appears in every component of $\vec{v}$ and $\vec{z}$ so it plays no role when $\vec{v}$ is linearly compared to $\vec{z}$. In the example, dropping $a_2$ would lead to $(v_1, v_2) = (a_1,a_3)$ and $(z_1, z_2) = (0, a_1+a_3)$. In this paper we choose to retain the reductive variable because it keeps the notation easier.
\end{enumerate}
The majorization test demands $\vec{v} \prec \vec{z}$ (read ``$\vec{z}$ majorizes $\vec{v}$\,''), namely:
\begin{itemize}
\item Setting variables $a_q$ ($q \neq p$) to arbitrary positive values, we demand  
\begin{equation}
\!\!\!\!\!\!\!\!\!
\max \big( v_{i_1} + v_{i_2} + \ldots + v_{i_k} \big)  \leq \max \big( z_{i_1} + z_{i_2} + \ldots + z_{i_k} \big) 
\quad \textrm{for all $1 \leq k \leq L_p$}~~
\label{defmaj}
\end{equation}
where the maxima are taken over all choices of $k$-tuples from components of $\vec{v}$ and $\vec{z}$.
\end{itemize}
If the initial inequality~(\ref{schematic}) is superbalanced then condition~(\ref{defmaj}) automatically holds with equality for $k = L_p = R_p$ because:
\begin{equation}
\sum_{i=1}^{L_p} v_i = \sum_{q=1}^N a_q L_{pq} 
\qquad {\rm and} \qquad 
\sum_{j=1}^{R_p} z_i = \sum_{q=1}^N a_q R_{pq}
\end{equation}

\subsection{New properties}
We prove the following:
\begin{itemize} 
\item Every null reduction of every \emph{balanced} holographic entropy inequality passes the majorization test. We show this in Section~\ref{major}.
\item Every null reduction of every \emph{superbalanced} holographic entropy inequality is a valid holographic entropy inequality. We show this in Section~\ref{valid}.
\end{itemize}
Both assertions were first conjectured in \cite{majorization}. They imply interesting corollaries, which we collect in Section~\ref{sec:corollaries}. The motivation that underlay the conjectures of \cite{majorization}, as well as the crisp form of the corollaries, leave much room for physical interpretations, including those concerning the dynamics of holographic theories of gravity. We postpone interpretive comments until the Discussion.

In this work, we assume that all holographic entropy inequalities can be proven by the contraction method \cite{hec}, which we briefly review in Section~\ref{revcontr}. We acknowledge the conceivable (in our view unlikely) possibility that some holographic entropy inequality may be valid yet admit no proof by the contraction method. References~\cite{allineqs, completecontract} argue that no such inequality exists.

\paragraph{Note added:} As we were finishing this manuscript, the authors of \cite{combinatorial} shared with us a draft of their paper, which derives the same results and disproves some prior conjectures; see Discussion for more details. We thank the authors of \cite{combinatorial} for synchronizing the arXiv submissions of our manuscripts.

\section{Contraction proof guarantees success on majorization test}
\label{major}

\subsection{A lightning review of contraction proofs}
\label{revcontr}
An efficient way to represent an inequality~(\ref{schematic}) is in terms of its indicator matrices $\mathcal{X}$ and $\mathcal{Y}$. We shall write the matrix elements of $\mathcal{X}$ as $x^p_i$ and those of $\mathcal{Y}$ as $y^p_j$. 

Matrix $\mathcal{X}$ is an $N \times L$ matrix of 0's and 1's, which tell us whether $A_p$ ($1 \leq p \leq N$) is contained in region $X_i$ ($1 \leq i \leq L$ where $L$ counts LHS terms in the inequality.) Matrix $\mathcal{Y}$ is defined analogously for the RHS; its dimensions are $N \times R$. Written in terms of matrix elements of $\mathcal{X}$ and $\mathcal{Y}$, inequality~(\ref{schematic}) takes the form:
\begin{equation}
\sum_{i=1}^L S\left( \cup_{p\, |\, x_i^p = 1} A_p\right) 
\geq  
\sum_{j=1}^R S\left( \cup_{p\, |\, y_j^p = 1} A_p\right) 
\end{equation}

As we shall see momentarily, the row vectors
\begin{equation}
\vec{x}\,^p \equiv \{x^p_i\}_{i=1}^N
\end{equation}
of matrix $\mathcal{X}$ play a special role. For obvious reasons, we call $\vec{x}\,^p$ the LHS indicator vector of $A_p$. Similarly, the row vectors of matrix $\mathcal{Y}$, denoted $\vec{y}\,^p$, are RHS indicator vectors of $A_p$. In what follows we denote components of vectors with subscripts, e.g. $(\vec{x}\,^p)_i = x^p_i$.

Consider the $L$-dimensional $\mathbb{F}_2$-vector space $\{0,1\}^L$ with the norm:
\begin{equation}
|| \vec{x} - \vec{x}\,' || = \sum_{i=1}^L | x_i - x'_i |
\end{equation}
Define a similar norm
\begin{equation}
|| \vec{y} - \vec{y}\,' || = \sum_{j=1}^R | y_j - y'_j |
\end{equation}
on the $R$-dimensional $\mathbb{F}_2$-vector space $\{0,1\}^R$. Reference~\cite{hec} showed that the validity of inequality~(\ref{schematic}) is ensured by the existence of a map $f: \{0, 1\}^L \to \{0, 1\}^R$, which has the following two properties:
\begin{itemize}
\item Contraction condition:
\begin{equation}
|| \vec{x} - \vec{x}\,' || \geq || f(\vec{x}) - f(\vec{x}\,') || \quad \forall\, \vec{x}, \vec{x}\,' \in \{0,1\}^L
\label{defcontr}
\end{equation}
\item Boundary conditions:
\begin{equation}
f(\vec{x}\,^p) = \vec{y}\,^p \quad {\rm for}~~0 \leq p \leq N 
\label{defbc}
\end{equation}
Here we supplement the $N$ indicator vectors $\vec{x}\,^p$ and $\vec{y}\,^p$ by an additional $(N+1)^{\rm st}$ pair $\vec{x}\,^0 \equiv \vec{0}$ and $\vec{y}\,^0 \equiv \vec{0}$. They are indicators of the environment region $A_0$, which is the complement of $A_1 A_2 A_3\!\ldots$
\end{itemize}

How conditions~(\ref{defcontr}) and (\ref{defbc}) prove inequality~(\ref{schematic}) has been reviewed in many references; see e.g. \cite{hec}. It is widely believed (see for example \cite{completecontract}) that all holographic entropy inequalities can be proven by constructing an appropriate contraction map. All arguments in this paper are based on the assumption that a contraction map, which proves inequality~(\ref{schematic}) exists. If some holographic entropy inequality is valid yet admits no proof by contraction, our arguments do not apply to it.

\subsection{Main argument}
\label{majormajor}
The ordering of the LHS terms in inequality~(\ref{schematic}) is arbitrary so we can reshuffle the summation index $i$ without consequence. To study the $p^{\rm th}$ null reduction of (\ref{schematic}), let us exploit that freedom and re-index the terms so that regions $X_i \supseteq A_p$ land in the first $L_p$ places and those without $A_p$ land in the last $L - L_p$ places. Then the components of vector $\vec{v}$ featured in condition~(\ref{defmaj}) can be written in the form:
\begin{equation}
v_i = \sum_{q=1}^N a_q x^q_i \qquad (1 \leq i \leq L_p)
\label{vjcomp}
\end{equation}
Let us denote the projection of vector $\vec{x}\,^q$ to the $A_p$-carrying components (the first $L_p$ components) as $\vec{x}\,^q|_p$ and the projection to the remaining coordinates as $\vec{x}\,^q|_{\not{p}}\,$:
\begin{equation}
\vec{x}\,^q = \vec{x}\,^q|_p + \vec{x}\,^q|_{\not{p}}
\label{xsplit}
\end{equation}
With a slight abuse of notation---ignoring a terminal string of $(L - L_p)$ zeroes, which are technically present in $\vec{x}\,^q|_p$---equation~(\ref{vjcomp}) is simply:
\begin{equation}
\vec{v} = \sum_{q =1}^N a_q\, \vec{x}\,^q|_p
\label{vdecomp}
\end{equation}
On the RHS, we set analogous definitions and reorder terms as necessary to arrive at:
\begin{equation}
\vec{z} = \sum_{q = 1}^N a_q\, \vec{y}\,^q|_p\,.
\label{zdecomp}
\end{equation}
We prove result~(\ref{defmaj}) in the following form:
\begin{itemize}
\item Set variables $a_q$ to arbitrary positive values and fix $1 \leq k \leq L_p$. For \emph{every} $k$-tuple of components of $\vec{v}$, say $\{v_{i_1}, v_{i_2} \ldots v_{i_k}\}$, there exists a $k$-tuple $\{z_{j_1}, z_{j_2} \ldots z_{j_k}\}$ of components of $\vec{z}$ such that:  
\begin{equation}
v_{i_1} + v_{i_2} + \ldots + v_{i_k}  \leq z_{i_1} + z_{i_2} + \ldots + z_{i_k}
\label{defmaj2}
\end{equation}
\end{itemize}
This statement is equivalent to (\ref{defmaj}). 

\paragraph{Inner products} 
To streamline the discussion, we define an inner product on $\{0, 1\}^L$ via:
\begin{equation}
|| \vec{x} - \vec{x}\,' || \equiv || \vec{x} || + || \vec{x}\,' || - 2 \langle \vec{x}, \vec{x}\,' \rangle
\end{equation}
In words, $\langle \vec{x}, \vec{x}\,' \rangle$ counts components, which are 1 both in $\vec{x}$ and in $\vec{x}\,'$. The inner product $\langle \vec{y}, \vec{y}\,' \rangle$ is defined analogously on $\{0, 1\}^R$; it also counts common 1's in $\vec{y}$ and in $\vec{y}\,'$. It is easy to see that these definitions satisfy the axioms of inner products.

Applied to the boundary conditions~(\ref{defbc}), these inner products efficiently encode balance and, when it holds, superbalance. For all $0 \leq p, q \leq N$ we have:
\begin{align}
{\bf balance\!:}
\qquad &
|| \vec{x}\,^p || = || \vec{y}\,^p || 
\label{super1} \\
{\bf superbalance\!:}
\qquad &
\langle \vec{x}\,^p, \vec{x}\,^q \rangle =  \langle \vec{y}\,^p, \vec{y}\,^q \rangle
\label{super2}
\end{align}

The inner products also afford a handy restatement of the majorization condition~(\ref{defmaj2}). Consider a vector $\vec{x}$ with $|| \vec{x} || = k$ and $\langle \vec{x}, \vec{x}\,^p \rangle = k$. This picks out $k$ LHS terms of (\ref{schematic}), which all contain $A_p$. We stipulate the existence of a vector $\vec{y}$ with $|| \vec{y} || = k$ and $\langle \vec{y}, \vec{y}\,^p \rangle = k$. Note that such a vector picks out $k$ RHS terms, which all contain $A_p$. 

In this language, claim~(\ref{defmaj2}) is synonymous with the following:
\begin{itemize}
\item Fix $p$ and $1 \leq k \leq L_p$. Given $\vec{x}$ with $|| \vec{x} || = k$ and $\langle \vec{x}, \vec{x}\,^p \rangle = k$, there exists a $\vec{y}$ with $|| \vec{y} || = k$ and $\langle \vec{y}, \vec{y}\,^p \rangle = k$ such that: 
\begin{equation}
\sum_{q =1}^N a_q \langle \vec{x}, \vec{x}\,^q \rangle 
\leq  
\sum_{q =1}^N a_q \langle \vec{y}, \vec{y}\,^q \rangle
\label{ipstatement}
\end{equation}
\end{itemize}
Comparing with equation~(\ref{vdecomp}), we see that it is alright to use $\vec{x}\,^q$ instead of $\vec{x}\,^q |_p$ because $\langle \vec{x}, \vec{x}\,^q|_{\not{p}} \rangle = 0$. Of course, the same comment applies to $\langle \vec{y}, \ldots \rangle$ on the RHS.

Since all $a_q$ are positive, it therefore suffices to prove:
\begin{itemize}
\item Fix $p$ and $1 \leq k \leq L_p$. Given $\vec{x}$ with $|| \vec{x} || = k$ and $\langle \vec{x}, \vec{x}\,^p \rangle = k$, there exists a $\vec{y}$ such that $|| \vec{y} || = k$ and $\langle \vec{y}, \vec{y}\,^p \rangle = k$ and:
\begin{equation}
\langle \vec{x}, \vec{x}\,^q \rangle \leq \langle \vec{y}, \vec{y}\,^q \rangle \qquad{\rm for}~1 \leq q \leq N
\label{ycond}
\end{equation}
\end{itemize}

\paragraph{Proof} We claim that if a contraction $f$ proves inequality~(\ref{schematic}) then $\vec{y} = f(\vec{x})$ satisfies the conditions in and above~(\ref{ycond}). 

It is useful to introduce the concept of {\bf colinearity}. For any three vectors $\vec{s}$, $\vec{t}$, $\vec{u}$ the triangle inequality holds:
\begin{equation}
|| \vec{u} - \vec{s} || \leq || \vec{u} - \vec{t} || + || \vec{t} - \vec{s} ||
\label{triangle}
\end{equation}
We say that $\vec{t}$ is \emph{colinear} with $\vec{s}$ and $\vec{u}$ if (\ref{triangle}) holds with equality. This concept is useful because of the following facts \cite{rg}:
\begin{itemize}
\item If $\vec{x}$ is colinear with $\vec{0} = \vec{x}\,^0$ and $\vec{x}\,^p$ then $f(\vec{x})$ is colinear with $\vec{0}$ and $f(\vec{x}\,^p) = \vec{y}\,^p$. 
\item $|| \vec{x} || = || f(\vec{x}) ||$ and $|| \vec{x}\,^p - \vec{x} || = || \vec{y}\,^p - f(\vec{x}) ||$.
\end{itemize}
The upper statement is true because
\begin{equation}
|| \vec{x}\,^p || = || \vec{x}\,^p - \vec{x} || + || \vec{x} || \geq || f(\vec{x}\,^p) - f(\vec{x}) || + || f(\vec{x}) || \geq || f(\vec{x}\,^p) || = || \vec{y}\,^p ||
\end{equation}
and balance in the form~(\ref{super1}) forces both $\geq$ signs to be equalities. The lower statement is true because now:
\begin{equation}
0 \geq || f(\vec{x}\,^p) - f(\vec{x}) || - || \vec{x}\,^p - \vec{x} || = || \vec{x} || - || f(\vec{x}) || \geq 0
\label{eq: novel_feature}
\end{equation}

In the problem at hand, the assumptions $|| \vec{x} || = k$ and $\langle \vec{x}, \vec{x}\,^p \rangle = k$ mean precisely that $\vec{x}$ is colinear with $\vec{0}$ and $\vec{x}\,^p$. Therefore, $|| f(\vec{x}) || = || \vec{x} || = k$ and
\begin{align}
2\langle f(\vec{x}), \vec{y}\,^p \rangle 
& = || f(\vec{x}) || + || \vec{y}\,^p || - || f(\vec{x}) - \vec{y}\,^p || 
\label{colinearlogic} 
\\ \nonumber
& = || \vec{x} || + || \vec{x}\,^p || - || \vec{x}\,^p - \vec{x} ||
= k + L_p - |L_p - k| = 2k
\end{align}
as desired. It remains to show that $\langle \vec{x}, \vec{x}\,^q \rangle \leq \langle f(\vec{x}), \vec{y}\,^q \rangle$ for $q \neq p$. This follows from:
\begin{align}
2 \langle \vec{x}, \vec{x}\,^q \rangle & = || \vec{x} || + || \vec{x}\,^q || - || \vec{x} - \vec{x}\,^q || \\
2 \langle f(\vec{x}), \vec{y}\,^q \rangle & = || f(\vec{x}) || + || \vec{y}\,^q || - || f(\vec{x}) - \vec{y}\,^q ||
\label{innerwrite2}
\end{align}
Since $|| \vec{x} || = k = || f(\vec{x}) ||$ and $|| \vec{x}\,^q || = || \vec{y}\,^q ||$ by (\ref{super1}), $\langle \vec{x}, \vec{x}\,^q \rangle \leq \langle f(\vec{x}), \vec{y}\,^q \rangle$ is simply the contraction condition.  

\subsection{Corollaries}
\label{sec:corollaries}

\paragraph{Main Corollary} Given a holographic entropy inequality proved by contraction $f$, take any $k$ LHS terms $\{X_{i_1}, X_{i_2} \ldots X_{i_k}\}$ with nonvanishing overlap $\cap_{m=1}^k X_{i_m} \neq \emptyset$. Then there exists a choice of $k$ RHS terms $\{Y_{j_1}, Y_{j_2} \ldots Y_{j_k}\}$ such that:
\begin{itemize}
\item The overlap of the $Y_{j_m}$'s is at least as large as the overlap of the $X_{i_m}$'s:
\begin{equation}
\cap_{m=1}^k X_{i_m} \subseteq \cap_{m=1}^k Y_{j_m}
\label{overlapclaim}
\end{equation}
\item Every atomic region $A_q$ appears in the collection $\{Y_{j_1}, Y_{j_2} \ldots Y_{j_k}\}$ at least as many times as it does in collection $\{X_{i_1}, X_{i_2} \ldots X_{i_k}\}$: 
\begin{equation}
\forall\, q: \quad \# \{m~|~ A_q \subseteq X_{i_m} \} \leq \# \{m~|~ A_q \subseteq Y_{j_m} \}
\label{countclaim}
\end{equation}
\end{itemize}
This corollary appears in paper~\cite{combinatorial} under the name ``Dominance.'' 
\medskip

To verify the corollary, consider vector $\vec{x}$ whose nonvanishing entries indicate the indices $i_m$ in collection $\{X_{i_1}, X_{i_2} \ldots X_{i_k}\}$. For every region $A_p \subseteq \cap_{m=1}^k X_{i_m}$, we see that $\vec{x}$ is colinear with $\vec{0}$ and $\vec{x}\,^p$. Therefore, $f(\vec{x})$ simultaneously satisfies conditions in and above (\ref{ycond}) for all $A_p \subseteq \cap_{m=1}^k X_{i_m}$. It follows that collection $\{Y_j~|~ f(\vec{x})_j = 1\}$ satisfies the claim of the corollary.

In particular, $\langle f(\vec{x}), \vec{y}\,^p \rangle = || f(\vec{x}) ||$ means that every region $Y_j$ with $f(\vec{x})_j = 1$ contains $A_p$. This establishes~(\ref{overlapclaim}). For claim~(\ref{countclaim}), we recognize that the count of the $X_{i_m}$'s which contain $A_q$ is $\langle \vec{x}, \vec{x}\,^q \rangle$ whereas the count of $Y_{j_m}$'s which contain $A_q$ is $\langle f(\vec{x}), \vec{y}\,^q \rangle$.

\paragraph{Remarks}
The $k=1$ instance of the Corollary is the statement that every $X_i$ is contained in at least one $Y_j$. Reference~\cite{rg} recently proved this special case as part of an argument, which interprets holographic entropy inequalities in terms of the holographic Renormalization Group. We return to this point in the Discussion.

The assumption that the $k$ chosen $X_i$'s must have at least one region in common is essential. For an example, take all the LHS regions in monogamy~(\ref{mmi}). Clearly no choice of $k=3$ RHS regions contains two copies of $A_1$, $A_2$ and $A_3$ each. In fact, with the single exception of subadditivity, we obtain a similar counterexample whenever we pick all the LHS regions in a maximally tight holographic entropy inequality. This observation leads to our next corollary.

\paragraph{Further corollaries} We list two: 
\begin{itemize}
\item Every maximally tight holographic entropy inequality has the property that its LHS terms have vanishing overlap: $\cap_{i=1}^L X_i = \emptyset$. 

The assertion is true for subadditivity. For facets other than subadditivity, Reference~\cite{sirui} conjectured and Reference~\cite{newineqs} later proved that each of them has strictly fewer LHS terms than RHS terms: $L < R$. For a contradiction, suppose that $\cap_{i=1}^L X_i \neq \emptyset$. Then the Main Corollary applies and one can choose $L$ terms on the RHS, which contain precisely $L_p$ copies of $A_p$ (for all $p$). In other words, these $L$ terms exhaust by themselves the budget of RHS appearances of every atomic region, which is afforded by the balance condition~(\ref{super1}). There is nothing from which to assemble a nonempty $(L+1)^{\rm st}$ RHS region whose existence is guaranteed by \cite{sirui, newineqs}.
\item We can slightly generalize the titular claim of Section~\ref{major}. Thus far, we have argued that ``single'' null reductions of holographic entropy inequalities pass the majorization test. Now consider ``multiply null-reduced'' inequalities: starting from a valid holographic inequality, only retain terms that contain the union $A_{p_1} A_{p_2} \ldots A_{p_s}$. Then every multiply null-reduced inequality passes the majorization test.

This claim is implied by the Main Corollary. We can also establish it directly by running the argument in Section~\ref{majormajor} over all vectors $\vec{x}$, which are simultaneously colinear with $\vec{0}$ and each $\vec{x}\,^{p_l}$ ($l = 1, 2\ldots s$). 
\end{itemize}

\section{Null reductions of valid superbalanced inequalities are valid}
\label{valid}
Concretely, the null reduction of a superbalanced inequality, which is proved by a contraction map $f$, is also provable by contraction. We show this by exhibiting a requisite contraction.

To set the notation, we are looking for a map $\underline{\vec{x}} \to f_p(\underline{\vec{x}})$, which proves the $p^{\rm th}$ null reduction. The components of $\underline{\vec{x}} \in \{0, 1\}^{L_p}$ correspond to LHS terms in the null reduction and, by the same token, to the $A_p$-carrying LHS terms in the original inequality. We denote with $(\underline{\vec{x}}, \vec{0})$ the $L$-dimensional vector, which is obtained by appending $L-L_p$ zeroes to $\underline{\vec{x}}$. 

\paragraph{The contraction}
Define $f_p$ by:
\begin{equation}
f\big( (\underline{\vec{x}}, \vec{0}) \big) = \big( f_p(\underline{\vec{x}}), \vec{0}\big)
\label{deffp}
\end{equation}
The $\vec{0}$ on the right hand side refers to the $R - R_p$ dimensions of $\{0, 1\}$, which are associated to the $A_p$-excluding RHS terms of the original inequality.

By construction, every nonzero entry in $(\underline{\vec{x}}, \vec{0})$ corresponds to an $A_p$-carrying LHS term. Therefore, $(\underline{\vec{x}}, \vec{0})$ is colinear with $\vec{0}$ and $\vec{x}\,^p$ in $\{0,1\}^L$. By the logic spelled out above equation~(\ref{colinearlogic}), this implies 
\begin{equation}
|| f\big( (\underline{\vec{x}}, \vec{0}) \big) || = \langle f\big( (\underline{\vec{x}}, \vec{0}) \big), \vec{y}\,^p \rangle 
\end{equation}
so \emph{every} nonvanishing component of $f\big( (\underline{\vec{x}}, \vec{0})\big)$ corresponds to an $A_p$-carrying RHS term. This ensures that $f\big( (\underline{\vec{x}}, \vec{0}) \big)$ takes the form $(\ldots, \vec{0})$ so equation~(\ref{deffp}) is well defined. It is also a contraction because:
\begin{equation}
|| \underline{\vec{x}} - \underline{\vec{x}}\,' ||_p = 
|| (\underline{\vec{x}}, \vec{0})  - (\underline{\vec{x}}\,', \vec{0}) || \geq 
|| f\big( (\underline{\vec{x}}, \vec{0})\big)  - f\big((\underline{\vec{x}}\,', \vec{0})\big) || =
|| f_p(\underline{\vec{x}})  - f_p(\underline{\vec{x}}\,') ||_p
\end{equation}
Here $|| \ldots ||_p$ is the norm on $\{0, 1\}^{L_p}$, which is to be employed in a contraction proof of the null reduction.

\paragraph{Boundary conditions} 
All that remains is to verify that $f_p$ respects the correct boundary conditions. Consider the LHS indicator vector $\underline{\vec{x}}\,^q$, which sets the $q^{\rm th}$ boundary condition for proving the $p^{\rm th}$ null reduction. In comparison with $\vec{x}\,^q|_p$ from equation~(\ref{xsplit}), it satisfies:
\begin{equation}
\vec{x}\,^q|_p = ( \underline{\vec{x}}\,^q, \vec{0})
\end{equation}
Of course, an analogous statement applies to the RHS:
\begin{equation}
\vec{y}\,^q|_p = ( \underline{\vec{y}}\,^q, \vec{0})
\end{equation}
Thus, the boundary condition $f_p(\underline{\vec{x}}\,^q) = \underline{\vec{y}}\,^q$ boils down to:
\begin{equation}
f(\vec{x}\,^q|_p) = \vec{y}\,^q|_p
\label{confbc}
\end{equation}
We observe that $\vec{x}\,^q|_p$ is collinear with $\vec{0}$ and $\vec{x}\,^q$, which implies:
\begin{equation}
|| f(\vec{x}\,^q|_p) || = || \vec{x}\,^q|_p || = \langle \vec{x}\,^q|_p, \vec{x}\,^q \rangle 
= \langle f(\vec{x}\,^q|_p), \vec{y}\,^q \rangle = \langle f(\vec{x}\,^q|_p), \vec{y}\,^q|_p \rangle
\label{longchain}
\end{equation} 
The first three equalities follow from collinearity by the argument around (\ref{colinearlogic}) and the last equality is implied by $f(\vec{x}\,^q|_p)$ having the form $(\ldots, \vec{0})$. Comparing the leftmost and rightmost expressions in equation~(\ref{longchain}), we see that only those components of $f(\vec{x}\,^q|_p)$ can be nonvanishing, which do not vanish in $\vec{y}\,^q|_p$. Equation~(\ref{confbc}) will follow if $f(\vec{x}\,^q|_p)$ has as many nonvanishing components as $\vec{y}\,^q|_p$ does. 

In a superbalanced inequality, this is true because:
\begin{equation}
|| f(\vec{x}\,^q|_p) || = || \vec{x}\,^q|_p || = L_{pq}
\qquad {\rm and} \qquad
|| \vec{y}\,^q|_p || = R_{pq}
\end{equation}
In a holographic entropy inequality which is not superbalanced, the conclusion does not hold. A counterexample is strong subadditivity~(\ref{ssa}), which has $L_{13} = 0$ LHS terms that contain $A_1 A_3$ but $R_{13} = 1$ RHS term that contains $A_1 A_3$. Indeed, the null reduction of (\ref{ssa}) on $A_1$ is $S(A_1 A_2) \geq S(A_1 A_2 A_3)$, which is not valid. Comparing with our argument, the culprit is the contraction proof of (\ref{ssa}), which does not obey $f(\vec{x}\,^3|_1) = \vec{y}\,^3|_1$ and therefore $f_1$ defined in (\ref{deffp}) fails the $A_3$-boundary condition.

\section{Discussion}
The original motivation for the present paper was the question: if we turn on time dependence and compute boundary entropies using the HRT proposal \cite{hrt}, is it possible to violate a holographic entropy inequality? Observing that regions selected from a common boundary lightcone in the CFT vacuum automatically saturate holographic inequalities, Reference~\cite{majorization} identified a proposition which---if true---would guarantee that a small state deformation in this configuration does not lead to a violation of the inequality. This property is the majorization test, which we reviewed in Section~\ref{sec:props}. By eliminating a large class of potential counterexamples, the result of Section~\ref{majormajor} adds evidence that the holographic entropy inequalities provable by contraction bound the set of entropies achievable by all \emph{maximin} surfaces, not only \emph{minimal} surfaces.

If time-dependent configurations can saturate but not violate holographic entropy inequalities, we may infer that the inequalities ``know about'' the dynamics of bulk gravity. Considering that the inequalities are essentially theorems about minimal cuts through weighted graphs, one wonders whether the protection of those graph-theoretic statements is somehow hard-wired into Einstein's equations. Assuming this speculation is correct, it is imperative to identify the physical principle, which might mediate between graph theory and gravitational dynamics.

References~\cite{ineqerror, rg} recently pointed out two physical principles, which are intimately related to holographic entropy inequalities. One is quantum erasure correction \cite{erasure}---the idea that a bulk domain $\mathcal{B}$ stores logical data protected against the erasure of boundary regions, which are separated from $\mathcal{B}$ by a Ryu-Takayanagi surface. Paper~\cite{ineqerror} showed that when a holographic entropy inequality is saturated (holds with equality), many such regions automatically become empty. In this way, holographic entropy inequalities define order parameters for the protection of logical data against certain erasures. It is not difficult to imagine that such a principle might be respected by gravitational time evolution.

A related result~\cite{rg} highlights a link with the holographic Renormalization Group \cite{holorg}---the rule of thumb that the additional (holographic) dimension of the AdS spacetime geometrizes CFT scales. The argument there compares entanglement wedges\footnote{Restricted to a spatial bulk slice, the entanglement wedge of boundary system $A$ is the bulk domain \emph{between} $A$ and the minimal surface that computes $S(A)$ in equation~(\ref{mincut}). Subregion duality \cite{rhodual, rhodual2} posits that the physics in the entanglement wedge of $A$ can be reconstructed from $\rho_A$ alone.} of boundary regions $X_i$ and $Y_j$, which appear in a holographic entropy inequality~(\ref{schematic}). Consistency of subregion duality \cite{rhodual, rhodual2} requires that if $X_i \subseteq Y_j$ then their entanglement wedges are similarly contained. By the $k=1$ instance of our Main Corollary, this implies that in every holographic entropy inequality, the wedges of RHS regions $E(Y_j)$ collectively cover more of the bulk spatial slice than do entanglement wedges of LHS regions $E(X_i)$. When the inequality is saturated, their coverage becomes equal. In this way, the inequalities define order parameters for infrared correlations, which are not mediated by the ultraviolet.

In their present form, the arguments in \cite{rg} and \cite{ineqerror} rely heavily on the assumption of time reversal symmetry. On the other hand, they appear closely related to the majorization property proven in Section~\ref{majormajor}. The RG argument, in particular, is built entirely on the $k=1$ instance of our Main Corollary: the greater coverage of the bulk by the entanglement wedges of RHS regions $Y_j$ manifests the infrared physics, which is accessible from the $Y_j$'s but not from the $X_i$'s \emph{because} every $X_i$ is subsumed in some $Y_j$. It might be difficult to lift the physical interpretations of holographic inequalities championed in \cite{rg} and \cite{ineqerror} directly to time-dependent settings. For immediate follow-up work, we propose a more readily feasible goal: to interpret the majorization property proven in this paper in the language of quantum erasure correction and/or the holographic Renormalization Group.

\paragraph{Note added:} When we were finishing this manuscript, the authors of \cite{combinatorial} shared with us their draft, which significantly overlaps with and in some respects exceeds the scope of this paper. In particular, \cite{combinatorial} also disproves by counterexample the converse statements of our main conclusions: 
\begin{itemize}
\item Not every inequality whose null reductions pass the majorization test is valid. 
\item Not every inequality, whose null reductions are valid holographic inequalities, is holographically valid.
\end{itemize}
The arXiv submissions of this paper and \cite{combinatorial} are synchronized. 

\acknowledgments
We thank Guglielmo Grimaldi, Matthew Headrick, Veronika Hubeny and Pavel Shteyner for sharing a draft of their paper \cite{combinatorial} and for synchronizing the arXiv submissions of our manuscripts. We thank Charlie Cummings, Ricardo Esp{\'i}ndola, Xiaoliang Qi, Sirui Shuai and Dachen Zhang for discussions. BC thanks the organizers of the workshops ``Discussions on Quantum Spacetime'' held at LMSI (Mumbai, India) and ``AI for Physics'' held at the Aspen Center for Physics while XLW thanks the organizers of the 20$^{\rm th}$ Asian Winter School on Strings, Particles and Cosmology held at IISER Bhopal (India), where part of this work was completed. This research was supported by an NSFC grant number 12042505. 

BC gives special thanks to Stella Christie, Vice Minister of Higher Education, Science, and Technology of the Republic of Indonesia, for the care and support given in the final stages of preparation of this manuscript.

\end{document}